\documentclass[letterpaper,english,reprint, aps]{revtex4-1}
\usepackage[T1]{fontenc}
\usepackage[latin9]{inputenc}
\usepackage{babel}
\usepackage{units}
\usepackage{amsmath}
\usepackage{amssymb}
\usepackage[unicode=true]
 {hyperref}



\begin{document}
\preprint{}
\title{Effective conductivity of the multidimensional chessboard}
\author{Clinton DeW. Van Siclen}
\email{cvansiclen@gmail.com}

\affiliation{1435 W 8750 N, Tetonia, Idaho 83452, USA}
\date{12 October 2020}
\begin{abstract}
An algebraic formula for the effective conductivity of a $d$-dimensional,
two component chessboard is proposed. The derivation relies on the
self-duality of the square bond lattice, the principle of universality,
and the analytical capabilities of the Walker Diffusion Method.
\end{abstract}
\maketitle

\section{\label{sec:level1}introduction \protect\lowercase{}}

There is long-standing interest in the derivation of exact expressions
for the effective conductivity of two-component systems. These of
course rely on the symmetries of the system. By consideration of an
electric field applied across a two-dimensional (2D) medium that is
a square array of identical circular inclusions imbedded in a matrix,
Keller \citep{Keller} showed that
\begin{equation}
\sigma\left(p,\alpha;q,\beta\right)\:\sigma\left(p,\beta;q,\alpha\right)=\alpha\,\beta\label{eq:1}
\end{equation}
where $\sigma\left(p,\alpha;q,\beta\right)$ is the effective conductivity
of the system $\left(p,\alpha;q,\beta\right)$, which is the medium
with inclusions of conductivity $\alpha$ comprising areal fraction
$p$, and matrix of conductivity $\beta$ comprising areal fraction
$q=1-p$. Mendelson \citep{Mendelson} subsequently showed that this
relation holds as well when the inclusions are randomly distributed
over the matrix. Note that Eq. (\ref{eq:1}) is unaffected by the
exchange $\alpha\leftrightarrow\beta$, which puts a condition on
the 2D systems to which it applies. That is, all conductivity values
$\beta/\alpha\geq0$ are allowed.

A derivation of the effective conductivity $\sigma_{dD}$ of the $d$-dimensional,
two-component chessboard has not appeared in the literature. The following
section derives $\sigma_{2D}$ by utilizing the self-duality property
of a 2D chessboard constructed as a square (conducting) bond network.
Section III applies an analytic expression of the WDM \citep{CVS99}
to construct a formula for $\sigma_{dD}$. Final comments are made
in Section IV.

\section{use of self-duality}

The \textit{dual} of a two-component, 2D square bond lattice is constructed
by crossing each $\alpha$ bond in the original system $\left(p,\alpha;q,\beta\right)$
with a $\alpha^{-1}$ bond and crossing each $\beta$ bond with a
$\beta^{-1}$ bond. The dual system $\left(p,\alpha^{-1};q,\beta^{-1}\right)$
is again a 2D square bond lattice. Of relevance here is the fact that
the conductivity of the original system equals the resistivity of
its dual; that is,
\begin{equation}
\sigma\left(p,\alpha;q,\beta\right)\:\sigma\left(p,\alpha^{-1};q,\beta^{-1}\right)=1.\label{eq:2}
\end{equation}
In the particular case of the 2D chessboard (constructed of conducting
bonds), this relation becomes
\begin{equation}
\sigma\left(\nicefrac{1}{2},\alpha;\nicefrac{1}{2},\beta\right)\:\sigma\left(\nicefrac{1}{2},\alpha^{-1};\nicefrac{1}{2},\beta^{-1}\right)=1\label{eq:3}
\end{equation}
or equivalently
\begin{equation}
\sigma\left(\nicefrac{1}{2},\alpha;\nicefrac{1}{2},\beta\right)\:\sigma\left(\nicefrac{1}{2},\beta;\nicefrac{1}{2},\alpha\right)=\alpha\beta\label{eq:4}
\end{equation}
giving the analytical result
\begin{equation}
\sigma\left(\nicefrac{1}{2},\alpha;\nicefrac{1}{2},\beta\right)=\left(\alpha\beta\right)^{1/2}.\label{eq:5}
\end{equation}

According to the principle of universality, any physical description
and properties of a structure in space are unaffected by how that
space is discretized. Here the structure is a two-component chessboard,
so that Eq. (\ref{eq:5}) gives the effective conductivity $\sigma_{2D}$
for the regular chessboard that is a square site (pixel) lattice,
as well.

Note that Eqs. (\ref{eq:2}) and (\ref{eq:4}) are broadly applicable
as they do not actually specify the distribution of the $\alpha$
and $\beta$ bonds in the 2D systems $\left(p,\alpha;q,\beta\right)$
and $\left(\nicefrac{1}{2},\alpha;\nicefrac{1}{2},\beta\right)$.
In fact, Eqs. (\ref{eq:1}) and (\ref{eq:2}) are identical, suggesting
a duality-based proof of Keller's and Mendelson's results for their
2D matrix/inclusion systems.

Note that Eq. (\ref{eq:5}) must also give the effective conductivity
for a chessboard comprised of triangular domains of the two components.
This follows from universality, which disregards the underlying lattice
type. As the triangular chessboard has no higher-dimension counterparts
it is not considered further.

Unfortunately there is no duality ``trick'' available in higher
dimensions, to apply to cubic and hypercubic bond lattices.

As an aside, note that Eq. (\ref{eq:5}) remains correct in the special
case of a \textit{random} distribution of $\alpha$ and $\beta$ bonds
comprising the square bond lattice. But universality does not make
the case that this relation holds true for other 2D lattice types.
This is because the percolation threshold $p_{c}$ of the square bond
lattice happens to be $\nicefrac{1}{2}$. Thus the ``universal structure''
created by the random distribution of $\alpha$ and $\beta$ bonds
in the 2D square bond lattice is the incipient infinite cluster of
$\alpha$ bonds and particular distribution of like-bond cluster sizes,
that characterize a system at the percolation threshold {[}\citealp{S=000026A}{]}.
Universality implies instead the relation
\begin{equation}
\sigma\left(p_{c},\alpha;(1-p_{c}),\beta\right)=\left(\alpha\beta\right)^{1/2}\label{eq:14}
\end{equation}
for all 2D lattice systems having a random distribution of $\alpha$
and $\beta$ bonds or sites. For lattice systems having phase fractions
$\left(1-p_{c}\right)\leq p_{c}$, Eq. (\ref{eq:14}) allows $\beta/\alpha\geq0$
and so satisfies Eqs. (\ref{eq:1}) and (\ref{eq:2}). For lattice
systems having phase fractions $\left(1-p_{c}\right)>p_{c}$, Eq.
(\ref{eq:14}) is restricted to $\beta/\alpha\leq1$. {[}A proof appears
in Sec. IV.{]}

In contrast, WDM calculations (of the sort described in Ref. \citep{CVSperc})
for the effective conductivity of the 2D square \textit{site} system
$\left(\nicefrac{1}{2},\alpha;\nicefrac{1}{2},\beta\right)$ in which
the $\alpha$ and $\beta$ sites are distributed randomly, find $\sigma\left(\nicefrac{1}{2},\alpha;\nicefrac{1}{2},\beta\right)<\alpha\,r^{1/2}$
where $r\equiv\beta/\alpha$ and $\beta\neq\alpha$. Significantly,
this system cannot be duplicated by a square bond system. In view
of the duality-based derivation of Eq. (\ref{eq:2}), it is clear
that Keller's relation is restricted to 2D systems that satisfy universality
(meaning, they are unaffected by the choice of lattice type).

\section{application of the wdm}

The ``site'' implementation of the WDM \citep{CVS99} is based on
the relation
\begin{equation}
\sigma=\left\langle \sigma(\mathbf{r})\right\rangle D_{w}\label{eq:6}
\end{equation}
between the effective conductivity $\sigma$ of a multicomponent conducting
medium and the (dimensionless) diffusion coefficient $D_{w}$ of a
walker diffusing through the medium according to particular rules.
The factor $\left\langle \sigma(\mathbf{r})\right\rangle $ is the
volume-average conductivity of the medium. In the case of a two-component
medium, $D_{w}$ is a functional of the conductivity ratio $\beta/\alpha$,
and has a numerical value that reflects the morphology of the two-component
system. That is, $D_{w}(p,\alpha;q,\beta)=D_{w}(p,\beta^{-1};q,\alpha^{-1})$.
Thus the effects of dimension $d$ on the effective conductivity of
the system are expressed through the walker diffusion coefficient
$D_{w}$.

In the particular case of a multidimensional chessboard, the function
$D_{w}$ must be symmetric in $\alpha$ and $\beta$, meaning that
$D_{w}(\nicefrac{1}{2},\alpha;\nicefrac{1}{2},\beta)=D_{w}(\nicefrac{1}{2},\beta;\nicefrac{1}{2},\alpha)$,
and must equal zero if either $\alpha$ or $\beta$ is zero. The simplest
expression that satisfies these constraints (and additionally, that
$D_{w}=1$ when $\alpha=\beta$) is
\begin{equation}
D_{w}^{(dD)}=\left[\frac{2\,(\alpha\beta)^{1/2}}{\alpha+\beta}\right]^{t}\label{eq:7}
\end{equation}
where the exponent $t$ is a function of the dimension $d$.

The relation between $t$ and $d$ is discovered by use of the equation
\begin{equation}
D_{w}^{(dD)}=\frac{2}{\alpha+\beta}\:\sigma_{dD}\label{eq:8}
\end{equation}
for dimensions $1$ and $2$. In the case of the 1D chessboard,
\begin{equation}
\sigma_{1D}=\frac{L}{R}=\frac{L}{\sum\frac{l_{i}}{\sigma_{i}}}=\frac{L}{\left\langle \sigma_{i}^{-1}\right\rangle \sum l_{i}}=\frac{2\,\alpha\beta}{\alpha+\beta}\label{eq:9}
\end{equation}
where $R$ is the resistance of an extended length $L=\sum l_{i}$
of the chessboard. Thus
\begin{equation}
D_{w}^{(1D)}=\frac{4\,\alpha\beta}{\left(\alpha+\beta\right)^{2}}.\label{eq:10}
\end{equation}
In the case of the 2D chessboard,
\begin{equation}
D_{w}^{(2D)}=\frac{2\,(\alpha\beta)^{1/2}}{\alpha+\beta}.\label{eq:11}
\end{equation}
Evidently the exponent $t=2/d$, giving the general results
\begin{equation}
D_{w}^{(dD)}=\left[D_{w}^{(1D)}\right]^{1/d}=\left[\frac{4\,r}{\left(1+r\right)^{2}}\right]^{1/d}\label{eq:12}
\end{equation}
where $r\equiv\beta/\alpha$, and
\begin{equation}
\sigma_{dD}=\frac{\alpha+\beta}{2}D_{w}^{(dD)}=\left(\frac{\alpha+\beta}{2}\right)^{1-2/d}\left(\alpha\beta\right)^{1/d}.\label{eq:13}
\end{equation}

Unfortunately, numerical calculations for $\sigma_{dD}$ are stymied
by the infinite resolution needed to accommodate the sharp corners
of the chessboard domains. Possibly a conducting 3D chessboard could
be fabricated (by 3D printing), and its effective conductivity measured
and compared to that predicted by Eq. (\ref{eq:13}).

\section{concluding remarks}

The formula for $D_{w}^{(dD)}$ given by Eq. (\ref{eq:7}), and then
Eq. (\ref{eq:12}), incorporates the symmetries and characteristics
of the chessboard morphology in all dimensions. This is achieved by
its expression as an exponential function, where the dimensional dependence
resides in the exponent.

Another two-component structure that occurs in all dimensions appears
when $\alpha$ and $\beta$ sites, with $\alpha>\beta$, are distributed
randomly in the proportion $p_{c}$ and $\left(1-p_{c}\right)$, respectively
{[}\citealp{CVSperc},\citealp{Straley}{]}. For these square, cubic,
and hypercubic systems, the effective conductivity $\sigma\left(p_{c},\alpha;(1-p_{c}),\beta\right)=\alpha\,r^{u}$
so
\begin{equation}
D_{w}^{(dD)}=\frac{\sigma_{dD}}{\left\langle \sigma(\mathbf{r})\right\rangle }=\frac{r^{u}}{p_{c}+\left(1-p_{c}\right)r}\label{eq:15}
\end{equation}
where $r\equiv\beta/\alpha\leq1$, and the value of the exponent $u$
depends on the dimension (for example $u_{2D}=\nicefrac{1}{2}$ and
$u_{3D}=\nicefrac{3}{4}$).

Note that Eq. (\ref{eq:15}) is easily derived in the manner of Eq.
(\ref{eq:7}). In this case, the particular constraint is that $D_{w}^{(dD)}$
equals zero when $\beta$ (that is, $r$) equals zero, since the $\beta$
phase fraction $\left(1-p_{c}\right)>p_{c}$ for all systems of dimension
$d>2$. And of course $D_{w}^{(dD)}=1$ when $r=1$. Then the observation
that the $\beta$ phase fraction $\left(1-p_{c}\right)\rightarrow1$
as dimension $d\rightarrow\infty$ leads directly to the results $\sigma_{dD}\rightarrow\beta$
and $u_{dD}\rightarrow1$, meaning that all $u_{dD}<1$. Another consequence
is that the ratio $r$ cannot be greater than $1$. Allowing the full
range $r\geq0$ is equivalent to allowing the exchange $r\leftrightarrow r^{-1}$
in these equations. In that case, $\sigma\left(p_{c},1;(1-p_{c}),r^{-1}\right)=r^{-u}$,
which implies $\sigma\left(p_{c},r;(1-p_{c}),1\right)=r^{1-u}$. But
the right-hand side of this last equation goes to zero as $r\rightarrow0$,
since all $u_{dD}<1$, while the left-hand side does not, since the
conducting phase fraction $(1-p_{c})>p_{c}$ for all systems of dimension
$d>2$.

In contrast, the 2D system has phase fractions $\left(1-p_{c}\right)<p_{c}$,
so that both $\alpha$ and $\beta$ must be non-zero in order that
the system conduct. Thus both ratios $\beta/\alpha\leq1$ and $\alpha/\beta\leq1$
are allowed (meaning $r\geq0$). Upon the exchange $r\leftrightarrow r^{-1}$,
the equation $\sigma\left(p_{c},1;(1-p_{c}),r\right)=r^{u}$ is transformed
to $\sigma\left(p_{c},r;(1-p_{c}),1\right)=r^{1-u}$. As $u_{2D}=\nicefrac{1}{2}$,
these effective conductivities are equal over the entire range $r\geq0$.

The functional similarity of Eqs. (\ref{eq:12}) and (\ref{eq:15})
suggests (not proves) that the former is the correct description of
walker diffusion over a multidimensional chessboard.
\begin{acknowledgments}
I thank Professor Robert ``Bob'' Smith (Department of Geological
Sciences) for arranging my access to the resources of the University
of Idaho Library (Moscow, Idaho).
\end{acknowledgments}

\end{document}